\documentclass[twocolumn,showpacs,amsmath,amssymb,aps,pra]{revtex4-1}
\usepackage{graphicx,color,hyperref}

\begin{document}

\title{Optical interference in view of the probability distribution
of photon detection}
\author{Toru Kawakubo}
\author{Katsuji Yamamoto}
\affiliation{Department of Nuclear Engineering, Kyoto University,
Kyoto 615-8540, Japan}
\date{\today}

\begin{abstract}
We investigate interference of optical fields
by examining the probability distribution of photon detection.
The usual description of interference patterns in terms of
superposition of classical mean fields with definite phases
is elucidated in quantum fashion.
Especially, for interference of two independent mixtures of number states
with Poissonian or sub-Poissonian statistics,
despite lack of intrinsic phases, it is found that
the joint probability has a distinct peak manifold
in the multi-dimensional space of the detector outcomes,
which is along the trajectory of the mean-field values
as the relative phase varies on the unit circle.
Then, an interference pattern should mostly appear
in each shot of measurement as a point in the peak manifold
with a randomly chosen relative phase.
On the other hand, for super-Poissonian sources the mean-field description
is likely invalidated with rather broad probability distributions.
\end{abstract}


\maketitle

\section{Introduction}
\label{sec:introduction}

Interference is often considered as a signature of superposition
in quantum systems. In particular, interference in many-body systems
as a macroscopic quantum effect has been attracting many interests.
In usual experiments, two fields originating in a common source
are subject to interfere, namely, each particle interferes with itself
\cite{Glauber1963,Mandel1965}.
On the other hand, in many-boson systems including lasers
\cite{Magyar1963,Pfleegor1967,Paul1986}
and atomic Bose-Einstein condensates (BECs) \cite{Andrews1997},
interference has been observed
even between independently prepared particles, especially as spatial fringes
in a single shot indicating the second-order coherence.
Such interference is often explained
in terms of the spontaneous symmetry breaking
for the relative phase between the independent sources,
which presumes nonvanishing expectation values of the field operators
or classical mean fields with definite phases.
However, the symmetry breaking seems problematic
in the absence of real mechanism.
In BECs, a U(1) symmetry is relevant for
the global phase rotation of atomic wavefunctions, the breakdown of which
relies on a nonphysical interaction \cite{Naraschewski1996,Leggett2006}.
In optical systems, a U(1) symmetry is also imposed
from lack of an absolute phase reference, which describes
the effective photon-number conservation in optical processes
\cite{Molmer1997,Sanders2003,Bartlett2007}. 

The interference pattern observed in a single shot of measurement
for independent sources under the U(1) symmetry
has been attributed to the back-action of particle detection on the systems,
which causes localization of the relative phase
\cite{Javanainen1996,Molmer1997,Sanders2003,Laloe2005}.
Another approach to the interference is to calculate
the correlation functions of the particle numbers measured by
the different detectors, which show the spatial modulation.
By evaluating the statistical moments of the Fourier components of
the spatial modulation up to the fourth order, the plane-wave interference
of atomic BECs is predicted in a single run with a random phase
\cite{Naraschewski1996,Iazzi2011}.
This analysis exploits the nature of the plane-wave mode functions.

In this paper, we investigate the interference of optical fields
comprehensively under various configurations for sources and detectors,
which is based on the probability theory of quantum measurement.
Specifically, we examine the probability distribution of the photon numbers
which are registered by the detectors,
rather than evaluating the correlation functions of the field intensities
as the averages over many runs of measurement.
This approach is hence of direct relevance
to see the interference pattern in each single shot.
Especially, for interference of two independent mixtures of number states,
despite lack of intrinsic phases due to the U(1) symmetry,
the joint probability in the multi-dimensional space of the photon counts
at the detectors may have a distinct manifold of sharp peaks
along the trajectory of the mean-field values
as the relative phase varies from $ - \pi $ to $ \pi $.
Then, the photon-number outcomes in each shot of measurement
should mostly be realized as a point in the peak manifold
with a randomly chosen relative phase, exhibiting an interference pattern.
Hence, the probability distribution of photon detection provides
a quantitative criterion to inspect whether an interference pattern
appears or not as described with the classical mean fields.
We examine the single-shot interference patterns
for U(1)-invariant source fields with a variety of photon-number statistics.
It will turn out that the mean-field description is applicable
to independent fields with Poissonian or sub-Poissonian statistics,
whereas for super-Poissonian statistics it is likely invalidated
with rather broad probability distributions.

The rest of this paper is presented as follows.
In Sec. \ref{sec:source}, the quantum states of two source fields
for interference are described under the U(1) symmetry
of phase transformation representing the photon-number superselection rule.
In Sec. \ref{sec:detection}, the photon detection
for interference is described.
A model of photon-detection system is presented.
Then, the joint probability of the photon counts
at the detectors is given in terms of the mode functions
and operators for the source fields.
Furthermore, relation among a variety of interference setups
is viewed as scaling for detectors and sources.
In Sec. \ref{sec:mean-field}, the usual description of interference
with classical mean fields is examined in the quantum viewpoint
by inspecting the probability distributions of the photon counts.
In Sec. \ref{sec:numerical-analysis}, a detailed numerical analysis
is presented to confirm the features of interference
which are examined in the preceding sections.
Section \ref{sec:conclusion} is devoted to conclusion.
A derivation of the joint probability of the photon counts is presented
in Appendix \ref{app:derivation}.

\section{Source fields under photon-number superselection}
\label{sec:source}

We consider a system of optical fields, where two sources are contained
for interference, either independent with lack of intrinsic phases
or correlated with a definite relative phase.
The positive-frequency field operator $ \hat{\psi}({\bf x}, t) $
is given generally in terms of the annihilation operators $ \hat{a}_l $
for a complete set of mode functions $ \phi_l $:
\begin{equation}
\hat{\psi}({\bf x},t) = \sum_l \hat{a}_l \phi_l ({\bf x},t) .
\label{eq:psi-op}
\end{equation}
Here, the time evolution of the free field is represented
in the mode functions $ \phi_l ({\bf x},t) $,
which may be determined in practice by expanding $ \hat{\psi} $
alternatively in terms of the plane-wave modes.
In order to describe an interference experiment, the mode functions
are chosen suitably to provide the two source fields as
\begin{equation}
\hat{a} \equiv \hat{a}_1 , \hat{b} \equiv \hat{a}_2 .
\end{equation}
For instance, in interference between two wave packets of light
the wavevector distributions are localized
around the central wavevectors of the respective sources.
In the following we assume for simplicity
that all the photons are populated in the two source modes ($ l = 1, 2 $),
while the other modes ($ l \geq 3 $) are in the vacuum states.
This treatment will be almost valid in usual interference experiments.

The quantum systems such as optical fields empirically obey
the superselection rule based on the conservation of particles (photons).
This is represented by the U(1) symmetry,
implying the absence of an absolute phase reference for the Bose fields
\cite{Bartlett2007}.
Henceforth, we consider the quantum description of interference
practically for the U(1)-invariant source fields.
The density matrix of each source $ \hat{\rho}_s $ ($ s = a, b $),
respecting the U(1) symmetry, is given by a photon-number distribution
$ p_s (N) $ for a mixture of the number states,
or a phase-invariant coherent-state representation
$ \mathcal{P}_s ( | \alpha | ) $ ($ \alpha \equiv r_s e^{i \phi_s} $)
\cite{Sanders2003}:
\begin{eqnarray}
\hat{\rho}_s &=& \sum_{N=0}^\infty p_s (N) |N \rangle \langle N|
= \int \frac{d^2 \alpha}{2 \pi} \mathcal{P}_s ( | \alpha | )
| \alpha \rangle \langle \alpha |
\nonumber \\
& \equiv & \int_{- \pi}^{\pi} \frac{d \phi_s}{2 \pi}
\int_0^\infty r_s d r_s \mathcal{P}_s ( r_s )
| r_s e^{i \phi_s} \rangle \langle r_s e^{i \phi_s} | .
\label{eq:rho-s}
\end{eqnarray}
For example, a Poissonian source $ \hat{\Pi}(| \alpha |) $
with a mean photon number $ \bar{N} = | \alpha |^2 $ is specified as
\begin{eqnarray}
p(N ; \hat{\Pi}) &=& e^{- \bar{N}} \frac{\bar{N}^N}{N!} ,
\label{eq:Poisson-p}
\\
\mathcal{P}(r ; \hat{\Pi})
&=& 2 \delta (  r^2 - \bar{N} ) ,
\label{eq:Poisson-P}
\\
\hat{\Pi}(| \alpha |) &=& \int_{- \pi}^{\pi} \frac{d \phi}{2 \pi}
| \sqrt{\bar{N}} e^{i \phi} \rangle \langle \sqrt{\bar{N}} e^{i \phi} | .
\label{eq:Poisson}
\end{eqnarray}
The state of two independent sources is then given by
\begin{equation}
\hat{\rho}_{a \otimes b} \equiv \hat{\rho}_a \otimes \hat{\rho}_b ,
\label{eq:rho-ind}
\end{equation}
with the uncorrelated random phases $ \phi_a $ and $ \phi_b $
under the U(1) symmetry.

On the other hand, the two fields may originate
in a common U(1)-invariant source.
By denoting the operators $ \hat{c}_1 $ and $ \hat{c}_2 $, respectively,
for the original source mode and the orthogonal auxiliary mode
in the vacuum, the operators for the two source fields may be given
in terms of a unitary transformation,
\begin{equation}
\left( \begin{array}{c} \hat{a} \\ \hat{b} \end{array} \right)
= \left( \begin{array}{cc} c & - s e^{-i \delta} \\
s e^{i \delta} & c \end{array} \right)
\left( \begin{array}{c} \hat{c}_1 \\ \hat{c}_2 \end{array} \right) ,
\label{eq:ab-common}
\end{equation}
where $ 0 \leq s, c \leq 1 $, $ s^2 + c^2 = 1 $,
and $ \delta $ is a certain given phase.
Then, an original number state $ | N \rangle_1 $ provides entangled sources
preserving the U(1) symmetry as
\begin{eqnarray}
| N \rangle_1 | 0 \rangle_2
&=& \frac{( c \hat{a}^\dagger + s e^{i \delta} \hat{b}^\dagger )^N}
{\sqrt{N!}} | 0 \rangle_1 | 0 \rangle_2
\nonumber \\
&=& \sum_{K=0}^N
\frac{\sqrt{N!}}{\sqrt{K!}\sqrt{(N-K)!}}
c^{K} s^{N-K} e^{i(N-K) \delta}
\nonumber \\
&{}& \times | K \rangle_a | N-K \rangle_b .
\label{eq:commonN}
\end{eqnarray}
A common Poissonian state $ \hat{\Pi}_1 (| \alpha |) $
($ \alpha \equiv | \alpha | e^{i \phi} $) also provides
\begin{eqnarray}
&{}& \int_{- \pi}^{\pi} \frac{d \phi}{2 \pi}
( | \alpha \rangle \langle \alpha | )_1
\otimes ( | 0 \rangle \langle 0 | )_2
\nonumber \\
&{}& = \int_{- \pi}^{\pi} \frac{d \phi}{2 \pi}
( | c \alpha \rangle \langle c \alpha | )_a
\otimes ( | s e^{i \delta} \alpha \rangle
\langle s e^{i \delta} \alpha | )_b ,
\label{eq:commonP}
\end{eqnarray}
where the resultant two sources share the original random phase $ \phi $,
and develop the definite relative phase $ \delta $.
The fields from a general common source $ \hat{\rho}^{\rm com}_1 $
are represented in terms of the states given in the above
with the original $ p(N) $ or U(1)-invariant $ \mathcal{P}(| \alpha |) $
for $ \hat{\rho}^{\rm com}_1 $.

Furthermore, if the two sources can refer to a certain frame system
for specifying their relative phase $ \delta $,
they may be represented in the U(1)-invariant form as
\begin{eqnarray}
\hat{\rho}_{ab} ( \delta )
& \equiv & \int_{- \pi}^{\pi} \frac{d \phi}{2 \pi}
\int_0^\infty r_a d r_a \mathcal{P}_a ( r_a )
\int_0^\infty r_b d r_b \mathcal{P}_b ( r_b )
\nonumber \\
&{}& \times | r_a e^{i ( \phi + \delta )} \rangle
\langle r_a e^{i ( \phi + \delta )} |
\otimes | r_b e^{i \phi} \rangle \langle r_b e^{i \phi} | .
\label{eq:ref-com}
\end{eqnarray}
The state of independent sources in Eq.\ (\ref{eq:rho-ind})
is then given formally as
\begin{equation}
\hat{\rho}_{a \otimes b}
= \int_{- \pi}^{\pi} \frac{d \delta}{2 \pi} \hat{\rho}_{ab} ( \delta ) ,
\end{equation}
which is the average over the random relative phase $ \delta $.

\section{Photon detection for interference}
\label{sec:detection}

\subsection{Photon detection and probability distribution}
\label{subsec:probability}

In optical interference experiments,
a commonly used photodetector records the number of photoelectrons
emitted from the detector surface during a time interval $ T $.
The time and surface integrated photon-flux operator
for the electron emission at some detector $m$ is given
\cite{Kelley1964,Cook1982,Bondurant1985,Vogel2006} by
\begin{equation}
\hat{I}_m = \eta_m \int_0^T dt \int_{S_m} dxdy
\hat{\psi}^\dagger ({\bf x},t) \hat{\psi} ({\bf x},t) ,
\label{eq:I_m}
\end{equation}
where $ \eta_m $ is the quantum efficiency, and the $ z $ axis is taken
normal to the detector surface $S_m$.  The bandwidth $ \Delta \omega $
of the incident radiation is assumed to be small enough compared
with the central frequency $ \omega_0 $.  The photon-flux operators
in Eq.\ (\ref{eq:I_m}) may be expressed
as bilinear forms of the mode operators,
\begin{equation}
\hat{I}_m = \sum_{ll'} R^{(m)}_{ll'} \hat{a}_l^\dagger \hat{a}_{l'} ,
\end{equation}
where the Hermitian matrices $ R^{(m)} $ are obtained
from Eq.\ (\ref{eq:I_m}) by substitution
$ \hat{\psi}^\dagger \hat{\psi} \to \phi_l^* \phi_{l'} $ as
\begin{equation}
R^{(m)}_{ll'} = \eta_m \int_0^T dt \int_{S_m} dxdy
\phi_l^* ({\bf x},t) \phi_{l'} ({\bf x},t) .
\label{eq:R_m}
\end{equation}

The joint probability of the photon counts
$ n_1, \dotsc , n_M \equiv {\bf n}_M $ registered by the $ M $ detectors,
which characterizes the statistics of interference, is given
\cite{Kelley1964,Cook1982,Bondurant1985,Vogel2006} by
\begin{eqnarray}
P(n_1, \dotsc , n_M) & \equiv & P({\bf n}_M)
\nonumber \\
&=& {\rm Tr} \left[ \hat{\rho} : \prod_{m=1}^M
\frac{( \hat{I}_m )^{n_m}}{n_m!} e^{-\hat{I}_m} : \right] ,
\label{eq:PM}
\end{eqnarray}
where $ \mathopen{:} \mbox{ } \mathclose{:} $ stands for normal ordering.
(A derivation is presented in Appendix \ref{app:derivation}.)
Then, the reduction relation follows as
\begin{equation}
P({\bf n}_{M-1}) = \sum_{n_M} P({\bf n}_M) .
\end{equation}
The joint probability is also additive for a combination of source states as
\begin{equation}
\hat{\rho} = \sum_{i=1}^K c_i \hat{\rho}_i
\rightarrow P({\bf n}_M) = \sum_{i=1}^K c_i P_i({\bf n}_M) ,
\label{eq:PMsum}
\end{equation}
where $ c_1 + \dotsb + c_K = 1 $ ($ 0 < c_i \leq 1 $).
The flux operators are given specifically as
\begin{equation}
\hat{I}_m = R^{(m)}_{aa} \hat{a}^\dagger \hat{a}
+ R^{(m)}_{bb} \hat{b}^\dagger \hat{b}
+ R^{(m)}_{ab} \hat{a}^\dagger \hat{b} + R^{(m)}_{ba}
\hat{b}^\dagger \hat{a} ,
\label{eq:I_m-ab}
\end{equation}
where $ R^{(m)}_{aa} , R^{(m)}_{bb} > 0 $, and
\begin{equation}
\\
R^{(m)}_{ab} = R^{(m) *}_{ba}
\equiv \xi_m e^{i \theta_m} \sqrt{R^{(m)}_{aa} R^{(m)}_{bb}} ,
\label{eq:Rab}
\end{equation}
with a certain phase $ \theta_m $ and $ 0 \leq \xi_m \leq 1 $,
representing the visibility of interference pattern, in accordance with
the Cauchy-Schwartz inequality for Eq.\ (\ref{eq:R_m}).
It should be noted that the terms involving the vacuum modes
($ l \geq 3 $) are dropped in $ \hat{I}_m $
since they provide null contributions to Eq.\ (\ref{eq:PM})
as the normal-ordered expectation values.
The maximal visibility parameter $ \xi_m = 1 $
may usually be obtained by a suitable detection system,
where the variation of the relative phase between the mode functions,
$ \arg [ \phi_a^* ({\bf x}) \phi_b ({\bf x}) ] \approx \theta_m $,
is negligible on the detector surface $ S_m $
(e.g., the sources being apart sufficiently from the detectors),
together with $ \phi_{a,b} ({\bf x},t)
\approx \phi_{a,b} ({\bf x}) e^{-i \omega_0 t} $
for the quasi-monochromatic modes ($ \Delta \omega \ll \omega_0 $).
In this situation, the detector matrices representing the flux operators
in Eq.\ (\ref{eq:R_m}) may be factorized approximately
in terms of the stationary mode functions as
\begin{equation}
R^{(m)}_{ll'} \approx \kappa_m \phi_l^* ({\bf x}_m) \phi_{l'} ({\bf x}_m) ,
\label{eq:R_m-1}
\end{equation}
where $ {\bf x}_m \in S_m $ and $ \kappa_m = \eta_m T |S_m| $
with the detection area $ |S_m| $. Then, the flux-operator is expressed as
\begin{equation}
\hat{I}_m = \kappa_m \hat{\Psi}_m^\dagger \hat{\Psi}_m ,
\label{eq:I_m-1}
\end{equation}
with a superposition of the mode operators,
\begin{equation}
\hat{\Psi}_m
= \phi_{ma} \hat{a} + \phi_{mb} \hat{b} ,
\label{eq:Psi_m}
\end{equation}
where $ \phi_{ma} \equiv \phi_a ({\bf x}_m) $
and $ \phi_{mb} \equiv \phi_b ({\bf x}_m) $.

\subsection{Scaling for detectors and sources}
\label{subsec:scaling}

For the two independent U(1)-invariant sources
$ \hat{\rho} = \hat{\rho}_{a \otimes b} $ in Eq.\ (\ref{eq:rho-ind}),
the mean photon number measured at each detector is given by
\begin{eqnarray}
\langle n_m \rangle
&=& {\rm Tr} [ \hat{\rho}_{a \otimes b} \hat{I}_m ]
\nonumber \\
&=& R^{(m)}_{aa} \bar{N}_a + R^{(m)}_{bb} \bar{N}_b
\nonumber \\
& \equiv & \langle n_m \rangle_a + \langle n_m \rangle_b ,
\label{eq:n_m}
\end{eqnarray}
with the mean photon numbers $ \bar{N}_s
= {\rm Tr} [ \hat{\rho}_s \hat{s}^\dagger \hat{s} ] $
initially contained in the sources.
The interference term with $ R^{(m)}_{ab} = R^{(m) *}_{ba} $,
indicating the so-called second-order coherence, disappears
in Eq.\ (\ref{eq:n_m}) on average over many runs of measurement.
It, however, will be seen later in detail that the interference fringes
may arise in each shot \cite{Magyar1963}
as the outcomes $ \{ n_m \} $ of photon detection,
which are obtained according to the joint probability in Eq.\ (\ref{eq:PM}).

We note in Eq.\ (\ref{eq:n_m}) that the coefficients
$ R^{(m)}_{aa} $ and $ R^{(m)}_{bb} $ indicate the probabilities
for each photon from the respective sources to fall into detector $ m $.
They may represent the resolution of interference.
Specifically, $ R^{(m)}_{ss} \propto |S_m| $ decreases as $ 1/M \to 0 $,
but keeping $ \langle n_m \rangle_s = R^{(m)}_{ss}\bar{N}_s \gg 1 $
for high accuracy statistics, when the photons are measured
by continuously distributed many small detectors ($ M \gg 1 $),
resulting in a fine spatial interference fringes.
In this sense, as seen in Eq.\ (\ref{eq:n_m}),
a change of $ R^{(m)}_{ss} $ (or resolution) for the detectors
may be viewed alternatively as an modification of the source statistics.
Here, consider scaling of the detector matrices $ R^{(m)} $
(by change of the detection efficiencies $ \kappa_m = \eta_m T |S_m| $),
\begin{align}
\tilde{R}^{(m)}(q) &= R^{(m)} / q &( q > 0 ) ,
\label{eq:scaling}
\end{align}
and define the binomial distribution
\begin{align}
B^{N}_{N'} (q) & \equiv \binom{N}{N'} q^{N'} (1-q)^{N-N'}
& (0 \leq N' \leq N) .
\label{eq:Bq}
\end{align}
In evaluating the joint probability, $ \langle \mathopen{:}
( \hat{I}_1 )^{k_1} \dotsm ( \hat{I}_M )^{k_M} \mathclose{:} \rangle $
contained in Eq.\ (\ref{eq:PM}) are calculated for a number state
$ | N_a , N_b \rangle $ with the normal-ordered expectation values
$ \langle (\hat{b}^\dagger)^{k_b} (\hat{a}^\dagger)^{k_a}
\hat{a}^{k_a} \hat{b}^{k_b} \rangle
= [N_a !/(N_a-k_a)!] \times [N_b !/(N_b-k_b)!] $
($ k_a + k_b = k_1 + \dotsb + k_M $),
which are multiplied by $ q^{k_a} q^{k_b} $ under the scaling.
Then, by considering the relation $ q^k [N!/(N-k)!]
= \sum_{N^\prime = k}^N B^{N}_{N^\prime} (q) [N^\prime !/(N^\prime - k)!] $,
the effects of this scaling can be included in the source statistics
without changing the calculations in Eq.\ (\ref{eq:PM}) as
\begin{equation}
\tilde{p}_s (N;q) = \sum_{N'=N}^\infty p_s (N') B^{N'}_{N} (q) ,
\label{eq:eff_s}
\end{equation}
which is also normalized as the original $ p_s (N) $.
Hence, the number statistics of the sources
may be replaced with the effective ones in Eq.\ (\ref{eq:eff_s})
by the scaling of $ q $, reproducing the same joint probability:
\begin{equation}
\{ \tilde{R}^{(m)}(q) , \tilde{p}_s (N;q) \} \to P({\bf n}_M) .
\label{eq:R-p-P}
\end{equation}
This may be viewed as a renormalization transformation
among the number statistics.
It indicates intimate relation for interference phenomena
in a variety of setups for detectors and sources.
According to Eq.\ (\ref{eq:eff_s}), the mean $ \tilde{\bar{N}}_s $ and
variance $ \tilde{V}_s $ for the effective statistics are given
in terms of the original ones as
\begin{eqnarray}
\tilde{\bar{N}}_s &=& q \bar{N}_s ,
\label{eq:N-eff}
\\
\tilde{V}_s &=& q^2 V_s + (1-q)q \bar{N}_s ,
\label{eq:V-eff}
\end{eqnarray}
preserving $ \tilde{R}^{(m)} \tilde{\bar{N}}_s = R^{(m)} \bar{N}_s $.
Then, for a sub-Poissonian distribution ($ V_s < \bar{N}_s $),
the effective one is still sub-Poissonian
($ \tilde{V}_s < \tilde{\bar{N}}_s $) as
\begin{equation}
\tilde{V}_s / \tilde{\bar{N}}_s = q ( V_s / \bar{N}_s ) + 1 - q .
\label{eq:V-N-eff}
\end{equation}
The Poissonian form is preserved under the renormalization
up to the scaling of the mean as $ \tilde{\bar{N}}_s = q \bar{N}_s $.
On the other hand, for a super-Poissonian distribution
the effective one is still super-Poissonian.

\section{Mean-field description}
\label{sec:mean-field}

The interference pattern is usually described
with superposition of classical mean fields.
We here consider quantum theoretical reasoning of this picture.
In the mean-field description, the source-mode operators are replaced
with c-number complex amplitudes
as $\hat{a} \to \alpha $ and $ \hat{b} \to \beta $.
That is, the physical quantities in normal ordering are evaluated
as the expectation values for coherent states $ | \alpha , \beta \rangle
\equiv | \alpha \rangle_a | \beta \rangle_b $ with definite phases.
Then, the mean photon-number count at each detector is given by
\begin{eqnarray}
\bar{n}_m ( \delta )
&=& \langle \alpha , \beta | \hat{I}_m | \alpha , \beta \rangle
\nonumber \\
&=& \langle n_m \rangle_a + \langle n_m \rangle_b
\nonumber \\
&{}& + 2 \xi_m \sqrt{\langle n_m \rangle_a \langle n_m \rangle_b}
\cos( \delta + \theta_m ) ,
\label{eq:mean}
\end{eqnarray}
where $ \delta = \arg \alpha - \arg \beta $,
$ | \alpha |^2 = \bar{N}_a $, $ | \beta |^2 = \bar{N}_b $,
$ \langle n_m \rangle_a = R^{(m)}_{aa} \bar{N}_a $
and $ \langle n_m \rangle_b = R^{(m)}_{bb} \bar{N}_b $ as given
in Eq.\ (\ref{eq:n_m}) for the two independent U(1)-invariant sources.
The set of $ \{ \bar{n}_m ( \delta ) \} $ with the definite relative phase
$ \delta $ exhibits the interference pattern
with the cosine term in Eq.\ (\ref{eq:mean}),
which oscillates with $ \theta_m $ depending on the detector location.
Specifically for the usual case with $ \xi_m = 1 $ of maximal visibility,
Eqs.\ (\ref{eq:R_m-1}), (\ref{eq:I_m-1}) and (\ref{eq:Psi_m})
for $ R^{(m)}_{ll'} $ and $ \hat{I}_m $ provide the interference pattern
in terms of the superposition of the macroscopic wavefunctions
$ \alpha \phi_a $ and $ \beta \phi_b $ of the two source modes:
\begin{eqnarray}
\bar{n}_m ( \delta ) = \kappa_m
\left| \alpha \phi_a ({\bf x}_m) + \beta \phi_b ({\bf x}_m) \right|^2 .
\label{eq:mean-1}
\end{eqnarray}
We note in Eqs.\ (\ref{eq:Rab}), (\ref{eq:n_m}), (\ref{eq:N-eff})
and (\ref{eq:mean}) that $ \xi_m $, $ \theta_m $,
$ \langle n_m \rangle_a $, $ \langle n_m \rangle_b $,
and hence the mean-field description $ \bar{n}_m ( \delta ) $
is invariant under the $ q $-scaling in Eq.\ (\ref{eq:R-p-P}).

The joint probability of photon detection in Eq.\ (\ref{eq:PM}) is calculated
with $ \langle \alpha , \beta | : ( \hat{I}_m )^k : | \alpha , \beta \rangle
= [ \bar{n}_m ( \delta ) ]^k $ ($ k \geq 0 $)
for the mean-field description as
\begin{eqnarray}
P({\bf n}_M ; \delta ) &=& \prod_{m=1}^M
\frac{[ \bar{n}_m ( \delta ) ]^{n_m}}{n_m!} e^{- \bar{n}_m ( \delta )}
\nonumber \\
& \equiv & \prod_{m=1}^M P( n_m ; \delta ) .
\label{eq:PM-mean}
\end{eqnarray}
This is the product of the Poisson distributions $ P( n_m ; \delta ) $
at the respective detectors.
Hence, the interference pattern appears for the outcomes
in each shot of measurement
as $ \{ n_m \} \approx \{ \bar{n}_m ( \delta ) \} $
with the shot noise $ \Delta n_m = \sqrt{\bar{n}_m} $.

The mean-field description is, however, not directly applicable
to the independent U(1)-invariant sources in Eq.\ (\ref{eq:rho-ind})
with $ {\rm Tr} [ \hat{\rho}_{a \otimes b} \hat{a}^\dagger \hat{b} ]
= {\rm Tr} [ \hat{\rho}_a \hat{a}^\dagger ]
\times {\rm Tr} [ \hat{\rho}_b \hat{b} ] = 0 $,
eliminating the cosine term in  Eq.\ (\ref{eq:mean}):
\begin{equation}
\langle n_m \rangle = \langle n_m \rangle_a + \langle n_m \rangle_b
\not= \bar{n}_m ( \delta ) .
\end{equation}
Nevertheless, by experiments and theoretical calculations
the interference fringes are observed for Poissonian sources
(laser fields \cite{Magyar1963}) and sub-Poissonian sources
(optical number states \cite{Molmer1997,Sanders2003} and BECs
\cite{Andrews1997,Naraschewski1996,Javanainen1996,Laloe2005,Iazzi2011}).
In the following, we examine the validity of the mean-field description
for a variety of source fields under the U(1) symmetry.

\subsection{Poissonian sources
and the peak manifold of the joint probability distribution}

We first consider as a prototype the case
of two independent Poissonian sources,
\begin{eqnarray}
\hat{\rho}_{a \otimes b} = \hat{\Pi}_a \otimes \hat{\Pi}_b
= \int_{- \pi}^{\pi} \frac{d \phi}{2 \pi}
\int_{- \pi}^{\pi} \frac{d \delta}{2 \pi}
| \alpha , \beta \rangle \langle \alpha , \beta | ,
\label{eq:PiPi}
\end{eqnarray}
where $ \arg \alpha = \phi + \delta $ and  $ \arg \beta = \phi $.
The joint probability of photon detection for these Poissonian sources
is given according to the additivity in Eq.\ (\ref{eq:PMsum}) by
\begin{equation}
P_{\hat{\Pi}}({\bf n}_M)
\equiv P({\bf n}_M ; \hat{\Pi}_a \otimes \hat{\Pi}_b )
= \int_{- \pi}^{\pi} \frac{d \delta}{2 \pi} P({\bf n}_M ; \delta ) .
\label{eq:PM-Poissonian}
\end{equation}
This is the average of the mean-field description $ P({\bf n}_M ; \delta ) $
in Eq.\ (\ref{eq:PM-mean})
over the intrinsically random and unknown relative phase $ \delta $
under the U(1) symmetry representing the photon-number superselection rule.
It is considered here that $ \bar{n}_m ( \delta ) $
for $ P({\bf n}_M ; \delta ) $ is independent of the overall phase $ \phi $.
Even though the second-order coherence is not observed manifestly
in Eq.\ (\ref{eq:n_m}), the form of the joint probability
such as $ P({\bf n}_M ; \hat{\Pi}_a \otimes \hat{\Pi}_b ) $
in Eq.\ (\ref{eq:PM-Poissonian}) indicates the interference effects
between the two independent fields, which may be observed
as the single-shot interference patterns and intensity correlations.
This is readily seen as follows.
Note that each $ P({\bf n}_M ; \delta ) $ in Eq.\ (\ref{eq:PM-mean})
for the coherent states with a relative phase $ \delta $
has a sharp peak at the point
$ \{ \bar{n}_m ( \delta ) \} $ ($ \bar{n}_m ( \delta ) \gg 1 $)
in the $ M $-dimensional space of the photon counts $ {\bf n}_M $.
Then, Eq.\ (\ref{eq:PM-Poissonian}) implies that there exists
the manifold of these peaks along the closed trajectory
of the mean-field values, practically providing
the support of $ P_{\hat{\Pi}}({\bf n}_M) $:
\begin{equation}
{\rm supp}[ P_{\hat{\Pi}}({\bf n}_M) ]
\approx \{ \bar{n}_m ( \delta ) \} ( \delta : - \pi \to \pi ) .
\label{eq:peaks}
\end{equation}
According to this specific form of the joint probability
$ P_{\hat{\Pi}}({\bf n}_M) $, the actual outcomes $ \{ n_m \} $
are mostly realized as a point in the peak manifold
with some relative phase $ \delta_1 $ randomly chosen a posteriori:
\begin{equation}
\{ n_m \} \approx \{ \bar{n}_m ( \delta_1 ) \} .
\end{equation}
Therefore, an interference pattern is exhibited
in each shot of measurement as described with the mean fields.

In comparison, the photon detection may be made sequentially
first for the source $ \hat{\rho}_a $, and then after a time interval
for the source $ \hat{\rho}_b $.
For these two entirely separate sources without any coherence,
interference by no means occurs between them.
The total joint probability for these subsequent measurements
is given by combining incoherently the individual joint probabilities
$ P_a $ for $ \hat{\rho}_a $ and $ P_b $ for $ \hat{\rho}_b $:
\begin{eqnarray}
P^{\rm incoh}({\bf n}_M)
= \sum_{{\bf n}^\prime_M} P_a ( {\bf n}^\prime_M )
P_b ( {\bf n}_M - {\bf n}^\prime_M ) .
\end{eqnarray}
This incoherent joint probability is calculated explicitly
for the pair of $ \hat{\Pi}_a $ and $ \hat{\Pi}_b $ as
\begin{equation}
P^{\rm incoh}_{\hat{\Pi}}({\bf n}_M)
\propto \prod_{m=1}^M
\frac{[ \langle n_m \rangle_a + \langle n_m \rangle_b ]^{n_m}}{n_m!} ,
\end{equation}
which merely has a single peak in the $ {\bf n}_M $ space at the point
$ \{ \langle n_m \rangle_a + \langle n_m \rangle_b \} $
without the interference term,
in contrast with $ P_{\hat{\Pi}}({\bf n}_M) $.

\begin{figure}[t]
\centering
\scalebox{1.1}{
\includegraphics*{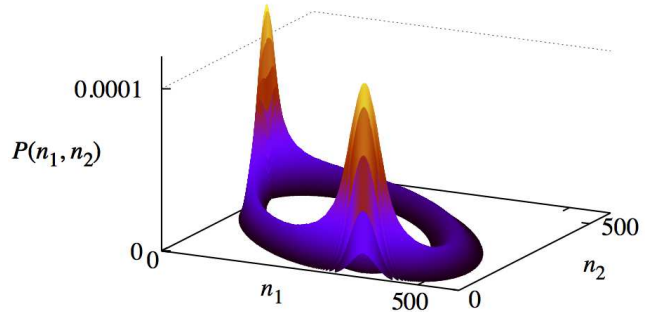}}
\caption{(Color online)
The joint probability $ P(n_1,n_2) $ is depicted
for the Poissonian sources $ \hat{\Pi}_a \otimes \hat{\Pi}_b $
with $ \bar{N}_a = \bar{N}_b = 500 $.
The peak manifold of $ P(n_1,n_2) $ exists along the mean-field trajectory
$ \{ \bar{n}_1 ( \delta ) , \bar{n}_2 ( \delta ) \}
( \delta : - \pi \to \pi ) $.
The conditional distribution $ P^\text{c}(n_2|n_1) $,
up to the normalization, provides a cross section of the peak manifold.}
\label{fig:P12}
\end{figure}
The feature of the joint probability $ P({\bf n}_M) $
for the Poissonian sources $ \hat{\Pi}_a \otimes \hat{\Pi}_b $
with $ \bar{N}_a = \bar{N}_b = 500 $,
as given in Eq.\ (\ref{eq:PM-Poissonian}),
is depicted on the $ (n_1,n_2) $ plane in Fig. \ref{fig:P12},
where the detector matrices $ R^{(m)} $ are taken typically
as in Eq.\ (\ref{eq:R-values}) (see Sec. \ref{sec:numerical-analysis}).
The peak manifold of $ P(n_1,n_2) $ appears clearly
along the mean-field trajectory
$ \{ \bar{n}_1 ( \delta ) , \bar{n}_2 ( \delta ) \}
( \delta : - \pi \to \pi ) $ as given in Eq.\ (\ref{eq:mean}).
In this case with $ \sqrt{\langle n_m \rangle_a}
\approx \sqrt{\langle n_m \rangle_b} $
($ R^{(m)}_{aa} \bar{N}_a \sim R^{(m)}_{bb} \bar{N}_b $)
and $ \xi_m = 1 $ for the maximal visibility ($ m = 1, 2 $),
the two prominent peaks are particularly seen in the regions
corresponding, respectively,
to $ \bar{n}_1 ( \delta_1 ) \ll \langle n_1 \rangle $
and $ \bar{n}_2 ( \delta_2 ) \ll \langle n_2 \rangle $
for certain $ \delta_1 $ and $ \delta_2 $
satisfying $ \cos( \delta_m + \theta_m ) = - 1 $.
The peak manifold becomes rather thin there,
and the probability distribution is squeezed
along the $ n_m $ direction with the narrow width
$ \sqrt{\bar{n}_m ( \delta_m )} $ of the Poisson distribution
$ P(n_m ; \delta_m ) $ for the specific $ \delta_m $.
That is, the peak is enhanced by a factor
$ \sim \sqrt{\langle n_m \rangle} / \sqrt{\bar{n}_m ( \delta_m )} $
(numerically about 3 here).
The probability for the interference pattern to be realized
within the area $ \sqrt{\langle n_1 \rangle} \sqrt{\langle n_2 \rangle} $
of the mean shot-noise level is, however, roughly uniform
along the peak manifold.
If there is a large difference between
the source photon numbers as $ \bar{N}_a \gg \bar{N}_b $
or $ \bar{N}_a \ll \bar{N}_b $, the mean-field trajectory
$ \{ \bar{n}_m ( \delta ) \} ( \delta : - \pi \to \pi ) $ shrinks
as $ 2 \sqrt{\langle n_m \rangle_a \langle n_m \rangle_b}
< \langle n_m \rangle_a + \langle n_m \rangle_b $ in Eq.\ (\ref{eq:mean}),
reducing effectively the visibility of the interference patterns.

\subsection{The conditional distributions
and estimation of the relative phase}

In order to check more closely the structure of the probability distribution
of photon detection for interference,
we note that despite lack of intrinsic phases due to the U(1) symmetry,
the outcomes at the detectors may provide estimation of the relative phase.
Specifically, we examine the conditional distributions of
the photon count at a detector given the outcomes at some other detectors.

The conditional distribution $ P^\text{c}(n_2|n_1) $
of the count $ n_2 $ at detector 2 with the outcome $ n_1 $ at detector 1
is given by
\begin{equation}
P^\text{c}(n_2|n_1) = \frac{P(n_1,n_2)}{P(n_1)} ,
\end{equation}
which provides a cross section of the peak manifold
of the joint probability $ P(n_1,n_2) $ in Fig. \ref{fig:P12},
up to the normalization with $ P(n_1) $.
By fitting the outcome $ n_1 $ to the mean-field value
$ \bar{n}_1 ( \delta ) $ in Eq.\ (\ref{eq:mean}),
an estimate of the relative phase $ \delta $ is obtained,
generally with two possibilities $ \delta^{\pm} $ due to the cosine.
Then, the outcome $ n_2 $ is inferred with the estimated phases as
\begin{equation}
n_1 = \bar{n}_1 ( \delta ) \rightarrow \delta^{\pm} (n_1)
\rightarrow \bar{n}_2^\pm \equiv \bar{n}_2 [ \delta^{\pm} (n_1) ] .
\end{equation}
Actually, if $ P^\text{c}(n_2|n_1) $ has the sufficiently narrow peaks
at $ \bar{n}_2^+ $ and $ \bar{n}_2^- $, the outcome $ n_2 $ should be
obtained almost at either of these peaks with high probability,
as predicted by the mean-field description.
The width of each peak should be
at most of the order of $ \sqrt{\bar{n}_2} $,
the shot noise level of the Poisson distribution
$ P( n_2 ; \delta ) = e^{- \bar{n}_2}(\bar{n}_2)^{n_2}/{n_2}! $,
in order to obtain the interference pattern.

Furthermore, we consider the conditional distribution of the count $ n_3 $
at detector 3 for the pair of the outcomes $ (n_1,n_2) $,
\begin{equation}
P^\text{c}(n_3|n_1,n_2) = \frac{P(n_1,n_2,n_3)}{P(n_1,n_2)} .
\end{equation}
Given the outcome at detector 1 as $ n_1 = \bar{n}_1 ( \delta^\pm) $,
the outcome at detector 2  will mostly be obtained
as either $ n_2 \approx \bar{n}_2^+ \equiv \bar{n}_2 [ \delta^+ (n_1) ] $
or $ n_2 \approx \bar{n}_2^- \equiv \bar{n}_2 [ \delta^- (n_1) ] $,
completing the estimation of the relative phase
$ \delta = \delta^+ $ or $ \delta = \delta^- $.
Then, if $ P^\text{c}(n_3| n_1 , n_2 ) $ of $ n_3 $
with $ n_1 = \bar{n}_1(\delta^\pm) $ and $ n_2 \approx \bar{n}_2^\pm $
has a single peak at $ \bar{n}_3^\pm \equiv \bar{n}_3[ \delta^\pm (n_1)] $
for either $ \delta^+ $ or $ \delta^- $,
the interference pattern mostly appears as $ \{ \bar{n}_1 ( \delta^+ ) ,
\bar{n}_2 ( \delta^+ ) , \bar{n}_3 ( \delta^+ ) \} $
or $ \{ \bar{n}_1 ( \delta^- ) , \bar{n}_2 ( \delta^- ) ,
\bar{n}_3 ( \delta^- ) \} $ according to the mean-field description.

Typically for the independent Poissonian sources
$ \hat{\Pi}_a \otimes \hat{\Pi}_b $ in Eq.\ (\ref{eq:PiPi}),
it is expected from Eq.\ (\ref{eq:PM-Poissonian})
that the conditional distributions behave as
\begin{eqnarray}
P^\text{c}_{\hat{\Pi}}(n_2|\bar{n}_1) \approx
C_+ P ( n_2 ; \delta^+ ) +  C_- P ( n_2 ; \delta^- ) ,
\label{eq:Pc2}
\end{eqnarray}
with certain normalization factors $ C_\pm \approx 1/2 $, and
\begin{eqnarray}
P^\text{c}_{\hat{\Pi}}(n_3|\bar{n}_1,\bar{n}_2^\pm )
\approx P( n_3 ; \delta^\pm ) .
\label{eq:Pc3}
\end{eqnarray}
These observations on the conditional distributions confirm
that the joint probability $ P_{\hat{\Pi}}({\bf n}_{\rm M}) $
has the peak manifold in Eq.\ (\ref{eq:peaks})
for the interference patterns,
as seen in Fig. \ref{fig:P12} for $ P_{\hat{\Pi}}(n_1,n_2) $.

\subsection{Sub-Poissonian sources}

We next argue that sub-Poissonian sources
lead to the narrower peaks in the probability distributions
than the Poissonian sources.
It is pointed out \cite{Pegg2009} that wave packets
emitted from a cavity maintain a pronounced relative phase coherence
when the intracavity field has a narrow number distribution.
Light beams from such sub-Poissonian cavities
will also exhibit the single-shot interference patterns
$ \{ n_m \} \approx \{ \bar{n}_m ( \delta ) \} $
($ - \pi \leq \delta < \pi $) as given by the mean-field description.
This phase coherence of each source is essential
to fix the interference phase between the independent sources
through the photon detection.

Specifically, consider two independent number states
\begin{equation}
\hat{\rho}_{a \otimes b} = \hat{N}_a \otimes \hat{N}_b
\equiv | N_a , N_b \rangle \langle N_a , N_b | .
\end{equation}
Since the Poissonian states $ \hat{\Pi}_a \otimes \hat{\Pi}_b $
are given in terms of the number states $ \hat{N}_a \otimes \hat{N}_b $
with the relevant photon-number distributions,
we have the relation between the joint probabilities
for these source states as
\begin{equation}
P_{\hat{\Pi}}({\bf n}_M)
= \sum_{N_a,N_b} p(N_a; \hat{\Pi}_a) p(N_b; \hat{\Pi}_b)
P_{\hat{N}}({\bf n}_M) .
\end{equation}
The Poissonian fluctuations of the photon numbers
in $ p(N_a; \hat{\Pi}_a) $ and $ p(N_b; \hat{\Pi}_b) $
will broaden the distribution of the outcomes $ {\bf n}_M $ to some extent
from $ P_{\hat{N}}({\bf n}_M) $.
Hence, it is inferred by consistency that $ P_{\hat{N}}({\bf n}_M) $
should also have the peak manifold of the mean-field description
$ \{ \bar{n}_m ( \delta ) \}(\delta : - \pi \to \pi) $ with
$ | \alpha |^2 = N_a $ and $ | \beta |^2 = N_b $ in Eq.\ (\ref{eq:mean}),
the width of which is narrower (smaller shot noise)
than the Poissonian $ P_{\hat{\Pi}}({\bf n}_M) $ as
\begin{align}
( \Delta n_m )_{\hat{N}} &= \gamma \sqrt{\bar{n}_m}
&( 0 < \gamma < 1 ) .
\label{eq:Dnm-N}
\end{align}

Generally, the independent fields $ \hat{\rho}_a \otimes \hat{\rho}_b $
are represented in terms of either the number states or the coherent states
in Eq.\ (\ref{eq:rho-s}).
The continuously distributed photon-number statistics of the sources
with variances $ \sim ( \Delta N )^2 $ broaden the probability distribution,
causing dispersion of the mean-field value $ \bar{n}_m (\delta) $
as a significant contribution to Eq.\ (\ref{eq:Dnm-N}) for the shot noise.
Since the mean-field value depends on the source photon numbers
roughly as $ \bar{n}_m \sim R \bar{N} $ for $ R^{(m)}_{ss} \sim R $
and $ \bar{N}_s \sim \bar{N} $ ($ s = a,b $), the shot noise,
or the width of the probability distribution of photon detection,
is estimated for $ \hat{\rho}_a \otimes \hat{\rho}_b $ as
\begin{equation}
( \Delta n_m )_{a \otimes b}
\sim \gamma \sqrt{\bar{n}_m} + ( \bar{n}_m / \bar{N} ) \Delta N .
\label{eq:Dnm}
\end{equation}
Particularly, $ ( \Delta n_m )_{\hat{\Pi}} = \sqrt{\bar{n}_m} $
implies $ 1 - \gamma \sim \sqrt{\bar{n}_m / \bar{N}} $
with $ \Delta N = \sqrt{\bar{N}} $.
Here, it should be remarked that the shot noise
for the interference of independent sources is not simply estimated
with the expectation values (statistical averages)
of the moments of the photon-flux operator $ \hat{I}_m $
($ \kappa_m \hat{\Psi}_m^\dagger \hat{\Psi}_m $).
This is due to the fact that the interference patterns vary run by run
with randomly chosen relative phases.

We also present an example of discrete photon-number distribution
for the sources, where the above estimate of the shot noise
in Eq.\ (\ref{eq:Dnm}) is not applicable simply.
That is, consider certain independent sources such as
\begin{eqnarray}
\hat{\rho}_a \otimes \hat{\rho}_b
&=& \frac{1}{4} ( \hat{N}^+_a + \hat{N}^-_a )
\otimes ( \hat{N}^+_b + \hat{N}^-_b )
\nonumber \\
&=& \frac{1}{4} \sum_{\sigma ,\sigma^\prime = \pm}
\hat{N}^{\sigma}_a \otimes \hat{N}^{\sigma^\prime}_b ,
\end{eqnarray}
which are the mixtures of two number states
\begin{eqnarray}
\hat{N}^\pm \equiv
| \bar{N} \pm \Delta N \rangle \langle \bar{N} \pm \Delta N | .
\end{eqnarray}
The mixed state $ ( \hat{N}^+ + \hat{N}^- )/2 $
is the U(1)-invariant form of $ ( | \bar{N} + \Delta N \rangle
+ | \bar{N} - \Delta N \rangle )/{\sqrt 2} $.
It has the variance of photon number $ V = ( \Delta N )^2 $.
In particular for the significant number difference
with $ \Delta N \sim \bar{N}/2 $, there are four peak manifolds
of $ \{ \bar{n}_m^{\pm \pm} ( \delta ) \} $, respectively,
for $ \hat{N}^\pm_a \otimes \hat{N}^\pm_b $
($ \sigma = \pm $, $ \sigma^\prime = \pm $),
which are mostly separated in the $ {\bf n}_M $ space.
Any one of these interference patterns appears randomly run by run
with the shot noise $ \sqrt{\bar{n}_m^{\pm \pm}} $.
Hence, the mixed state $ ( \hat{N}^+ + \hat{N}^- )/2 $ is regarded
as sub-Poissonian in spite of the apparent large variance
$ ( \Delta N )^2 \sim ( \bar{N}/2 )^2 $
since its photon-number distribution is essentially narrow.
The peak manifolds $ \{ \bar{n}_m^{++} ( \delta ) \} $
and $ \{ \bar{n}_m^{--} ( \delta ) \} $ are similar to each other,
providing the same interference pattern with contrast in brightness
by the factor $ \bar{n}_m^{++} / \bar{n}_m^{--}
= ( \bar{N} + \Delta N )/( \bar{N} - \Delta N ) $.
On the other hand, the peak manifolds $ \{ \bar{n}_m^{+-} ( \delta ) \} $
and $ \{ \bar{n}_m^{-+} ( \delta ) \} $ represent
the same interference pattern for $ R^{(m)}_{aa} = R^{(m)}_{bb} $.

\subsection{Super-Poissonian sources}

As for super-Poissonian sources, they are given relevantly
in the coherent-state representation with U(1)-invariant non-singular
$ \mathcal{P} $ functions.
Then, the joint probability of photon detection is given as
\begin{eqnarray}
P_{{\rm sup}\hat{\Pi}}({\bf n}_M)
= \int_{- \pi}^{\pi} \frac{d \delta}{2 \pi}
P_{{\rm sup}\hat{\Pi}}({\bf n}_M ; \delta ) ,
\end{eqnarray}
where
\begin{eqnarray}
P_{{\rm sup}\hat{\Pi}}({\bf n}_M ; \delta )
&=& \int_0^\infty \frac{d r_a^2}{2} \mathcal{P}_a (r_a)
\int_0^\infty \frac{d r_b^2}{2} \mathcal{P}_b (r_b)
\nonumber \\
&{}& \times P( {\bf n}_M; \delta ; r_a, r_b ) ,
\label{eq:PsupP}
\end{eqnarray}
with $ | \alpha | = r_a $ and $ | \beta | = r_b $
for $ \bar{n}_m ( \delta ; r_a, r_b ) $ in Eq.\ (\ref{eq:mean}).
Then, typically for thermal fields
$ \hat{\Theta}_a \otimes \hat{\Theta}_b $ with
$ \mathcal{P}_s (r_s; \hat{\Theta}_s) \propto \exp (- r_s^2 / \bar{N}_s) $,
the peak of $ P({\bf n}_M; \delta ) $ is smeared out in Eq.\ (\ref{eq:PsupP})
for each $ \delta $, providing the large shot noise
$ ( \Delta n_m )_{\hat{\Theta}} \sim \bar{n}_m $
with $ \Delta N \approx \bar{N} $ in Eq.\ (\ref{eq:Dnm}).
Hence, the interference pattern does not arise anyway
for independent super-Poissonian sources such as thermal states,
even though the outcomes after many runs of measurement
may manifest the higher-order coherence effects,
including the correlation of the photon counts.

\subsection{Correlated sources with a definite relative phase}

The usual interference pattern $ \{ \bar{n}_m ( \delta ) \} $
with the superposition of the classical mean fields,
as given in Eqs.\ (\ref{eq:mean}) and (\ref{eq:mean-1}),
arises for the pair of fields with a definite relative phase,
which originate in a common U(1)-invariant source $ \hat{\rho}^{\rm com}_1 $
through a unitary transformation in Eq.\ (\ref{eq:ab-common}).
This is understood by noting the relation from Eq.\ (\ref{eq:Psi_m}),
\begin{eqnarray}
\hat{\Psi}_m^\dagger \hat{\Psi}_m
&=& ( \phi_{ma}^* \hat{a}^\dagger + \phi_{mb}^* \hat{b}^\dagger )
( \phi_{ma} \hat{a} + \phi_{mb} \hat{b} )
\nonumber \\
&=& \left| c \phi_{ma} + e^{i \delta} s \phi_{mb} \right|^2
\hat{c}_1^\dagger \hat{c}_1 + \dotsb ,
\end{eqnarray}
up to the irrelevant terms involving the vacuum mode $ \hat{c}_2 $.
The mean photon count at each detector is given by
\begin{equation}
\langle n_m \rangle_{\rm com}
= {\rm Tr} [ \hat{\rho}^{\rm com}_1 \otimes ( | 0 \rangle \langle 0 | )_2
\hat{I}_m ]
= \bar{n}_m ( \delta ) ,
\label{eq:n_m-com}
\end{equation}
where $ \alpha = c \sqrt{\bar{N}} $
and $ \beta = e^{i \delta} s \sqrt{\bar{N}} $ in Eq.\ (\ref{eq:mean-1})
with the mean photon number of the common source
$ \bar{N}={\rm Tr}[ \hat{\rho}^{\rm com}_1 \hat{c}_1^\dagger \hat{c}_1 ] $.
The shot noise $ \Delta n_m $ is also given \cite{Mandel1965} by
\begin{eqnarray}
( \Delta n_m )^2
&=& \langle : ( \hat{I}_m )^2 : \rangle + \langle \hat{I}_m \rangle
- ( \langle \hat{I}_m \rangle )^2
\nonumber \\
&=& ( \bar{n}_m )^2 ( \Delta N / \bar{N} )^2
+ \bar{n}_m [ 1 - ( \bar{n}_m / \bar{N} ) ] ,
\end{eqnarray}
which is determined by the statistics of the common source
$ \hat{\rho}^{\rm com}_1 $.
Hence, for a super-Poissonian $ \hat{\rho}^{\rm com}_1 $
with $ \Delta N \sim \bar{N} $ such as a thermal state,
the interference pattern is not observed practically
in a single run of measurement due to the large shot noise.
It may rather appear by accumulating the outcomes of many runs
keeping the definite relative phase.
This is in contrast with the case of independent super-Poissonian sources,
where the interference pattern does not arise anyway
due to the random relative phases run by run.

The mean-field description for the interference pattern
$ \{ \bar{n}_m ( \delta ) \} $ in Eq.\ (\ref{eq:mean})
is similarly applicable to the case of the two source fields
$ \hat{\rho}_{ab} ( \delta ) $ with a definite relative phase $ \delta $
in Eq.\ (\ref{eq:ref-com}) by sharing the common reference frame.
The shot noise is determined depending on the source field statistics.

\section{Numerical analysis}
\label{sec:numerical-analysis}

We here present detailed numerical calculations
on the probability distributions of the photon counts
for a variety of independent U(1)-invariant source fields.
This analysis confirms the features of the single-shot interference
in terms of the mean-field description, which have been examined so far.
We show specifically the behavior of the joint probabilities
$ P(n_1,n_2) $ and $ P(n_1,n_2,n_3) $ for two and three detectors,
together with their conditional distributions
$ P^\text{c}(n_2|n_1)$ and $ P^\text{c}(n_3|n_1,n_2)$.

\subsection{Poissonian sources}
\label{sec:na-P}

As the prototype, we first present the results
for the Poissonian sources $ \hat{\Pi}_a \otimes \hat{\Pi}_b $,
which provide the essential understandings
how an interference pattern appears in each shot
according to the probability distribution of photon detection.
The detector matrices $ R^{(m)} $ ($ m=1,2,3 $)
in Eqs.\ (\ref{eq:I_m-ab}) and (\ref{eq:Rab}) are chosen typically as
\begin{align}
& R^{(1)}_{aa} = 0.3 , & & R^{(1)}_{bb} = 0.2 , & & \xi_1 = 1 , &
& \theta_1 = 0 ;
\nonumber \\
& R^{(2)}_{aa} = 0.2 , & & R^{(2)}_{bb} = 0.3 , & & \xi_2 = 1 , &
& \theta_2 = 0.7 \pi ;
\nonumber \\
& R^{(3)}_{aa} = 0.2 , & & R^{(3)}_{bb} = 0.3 , & & \xi_3 = 1 , &
& \theta_3 = - 0.5 \pi .
\label{eq:R-values}
\end{align}
The mean photon numbers of the Poissonian sources
$ \hat{\Pi}_a \otimes \hat{\Pi}_b $ are taken as
\begin{equation}
\bar{N}_a= \bar{N}_b = 500 ,
\label{eq:N-values}
\end{equation}
which provide the mean photon counts at the detectors in Eq.\ (\ref{eq:n_m}),
\begin{equation}
\{ \langle n_1 \rangle , \langle n_2 \rangle , \langle n_3 \rangle \}
= \{ 250 , 250 , 250 \} .
\end{equation}

\begin{figure}[t]
\centering
\scalebox{1.0}{
\includegraphics*{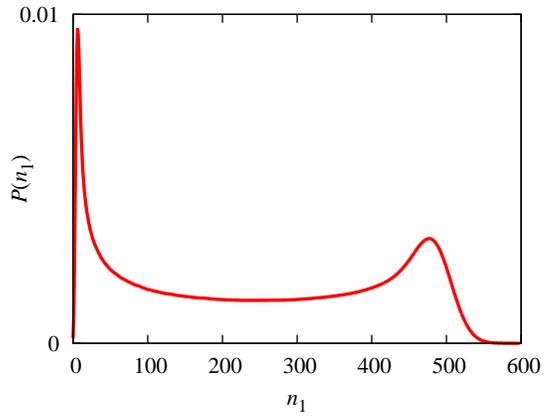}}
\caption{(Color online)
The probability distribution $ P(n_1) $ of the photon count $ n_1 $
at detector 1 is plotted for the Poissonian sources
$ \hat{\Pi}_a \otimes \hat{\Pi}_b $ with $ \bar{N}_a= \bar{N}_b = 500 $.
This distribution appears roughly flat
for $ 0 \lesssim n_1 \lesssim 2 \langle n_1 \rangle = 500 $,
corresponding to the range of $ \bar{n}_1 ( \delta ) $
for $ - \pi \leq \delta < \pi $,
as the Poisson distribution $ P(n_1; \delta ) $
in the mean-field description is averaged
over the unknown relative phase $ \delta $.}
\label{fig:P1}
\end{figure}

The probability distribution $ P(n_1) $
of the photon count $ n_1 $ at detector 1 is plotted in Fig. \ref{fig:P1},
which is given by
\begin{equation}
P(n_1) = \sum_{n_2 \dotsb n_M} P({\bf n}_M)
= \int_{- \pi}^{\pi} \frac{d \delta}{2 \pi} P( n_1 ; \delta ) .
\end{equation}
This distribution appears roughly flat
for $ 0 \lesssim n_1 \lesssim 2 \langle n_1 \rangle = 500 $,
corresponding to the range of $ \bar{n}_1 ( \delta ) $
for $ - \pi \leq \delta < \pi $,
as the Poisson distribution $ P(n_1; \delta ) $
in the mean-field description is averaged
over the intrinsically unknown relative phase $ \delta $.
It represents roughly the overview of the peak manifold
of $ P({\bf n}_M) $ along the $ n_1 $ axis.
The squeezed peak of $ P(n_1,n_2) $ corresponding to
$ \bar{n}_2 ( \delta_2 ) \ll \langle n_2 \rangle $ in Fig. \ref{fig:P12}
is smoothed out for $ P(n_1) $ by taking the sum over $ n_2 $,
while that corresponding to
$ \bar{n}_1 ( \delta_1 ) \ll \langle n_1 \rangle $ is still seen
around $ n_1 \approx 0 $.
It is also noticed that $ P(n_1) $ is enhanced around
$ n_1 \approx \bar{n}_1 ( \delta )_{\rm max} $
($ \approx 2 \langle n_1 \rangle = 500 $).
This is because the peak manifold of $ P(n_1,n_2) $
appears somewhat thicker there along the $ n_2 $ axis,
which is tangent to the mean-field trajectory.

\begin{figure}[t]
\centering
\scalebox{1.0}{
\includegraphics*{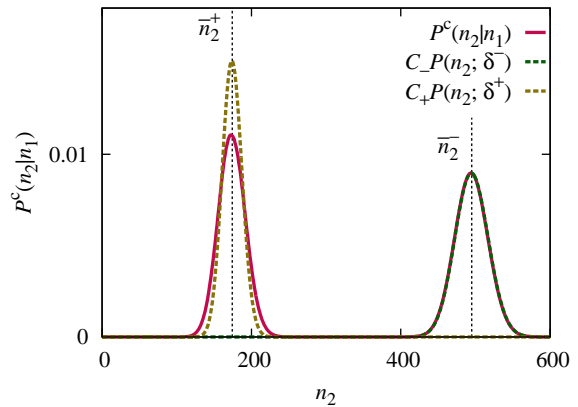}}
\caption{(Color online)
The conditional distribution $ P^\text{c}(n_2|n_1) $ (the solid line)
is plotted for the Poissonian sources $ \hat{\Pi}_a \otimes \hat{\Pi}_b $
with $ \bar{N}_a= \bar{N}_b = 500 $.
It is compared with the relevant Poisson distributions
$ C_+ P( n_2 ; \delta^+ ) $ (the left dashed line)
and $ C_- P( n_2 ; \delta^- ) $ (the right dashed line)
with $ C_\pm = 1/2 $, as suggested in Eq.\ (\ref{eq:Pc2}).
The mean-field values $ \bar{n}_2^+ \approx 174 $
and $ \bar{n}_2^- \approx 495 $ for $ n_1 = 106 $
are indicated with the vertical dotted lines.}
\label{fig:Pc2}
\end{figure}

The conditional distribution $ P^\text{c}(n_2|n_1) $
is plotted in Fig. \ref{fig:Pc2}.
Here, the first outcome is set, for example,
as $ n_1 = \bar{n}_1 ( \delta ) = 106 $,
which corresponds to the mean-field values
$ \bar{n}_2^+ \approx 174 $ with $ \delta^+ (n_1) = + 0.7 \pi $
and $ \bar{n}_2^- \approx 495 $
with $ \delta^- (n_1) = - 0.7 \pi $,
as indicated with the vertical dotted lines.
The relevant Poisson distributions
$ C_+ P( n_2 ; \delta^+ ) $ and $ C_- P( n_2 ; \delta^- ) $
with $ C_\pm = 1/2 $ are shown together,
as suggested in Eq.\ (\ref{eq:Pc2}).
The distribution around $ \bar{n}_2^- \approx 495 $ is in good agreement
with $ P( n_2 ; \delta^- )/2 $,
while that around $ \bar{n}_2^+ \approx 174 $ is slightly broader
than $ P( n_2 ; \delta^+ )/2 $.
This is due to the uncertainty in the estimation of
the relative phase $ \delta $ with the smaller $ \bar{n}_2^+ $.

\begin{figure}[t]
\centering
\scalebox{1.0}{
\includegraphics*{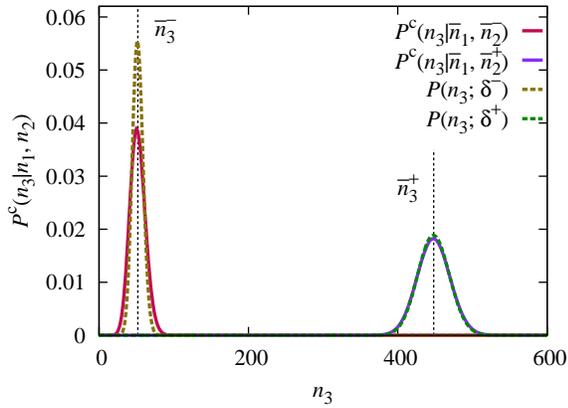}}
\caption{(Color online)
The conditional distributions $ P^\text{c}(n_3|n_1,n_2) $ (the solid lines)
are plotted for the Poissonian sources $ \hat{\Pi}_a \otimes \hat{\Pi}_b $
with $ \bar{N}_a= \bar{N}_b = 500 $,
where the mean-field values are taken as the outcomes,
$ (n_1,n_2) = ( \bar{n}_1 = 106 , \bar{n}_2^+ \approx 174 ) $
and $ ( \bar{n}_1 = 106 , \bar{n}_2^- \approx 495 ) $,
respectively, for $ \delta^+ = + 0.7 \pi $
and $ \delta^- = - 0.7 \pi $.
They are compared with the relevant Poisson distributions
$ P( n_3 ; \delta^+ ) $ (the right dashed line)
and $ P( n_3 ; \delta^- ) $ (the left dashed line),
as suggested in Eq.\ (\ref{eq:Pc3}).
The mean-field values $ \bar{n}_3^+ \approx 448 $
and $ \bar{n}_3^- \approx 52 $
are indicated with the vertical dotted lines.}
\label{fig:Pc3}
\end{figure}

The conditional distributions $ P^\text{c}(n_3|n_1,n_2) $
are plotted in Fig. \ref{fig:Pc3},
where the mean-field values are taken as the outcomes,
$ (n_1,n_2) = ( \bar{n}_1 = 106 , \bar{n}_2^+ \approx 174 ) $
and $ ( \bar{n}_1 = 106 , \bar{n}_2^- \approx 495 ) $,
respectively, for $ \delta^+ = + 0.7 \pi $
and $ \delta^- = - 0.7 \pi $.
They essentially agree with the relevant Poisson distributions
$ P( n_3 ; \delta^+ ) $ and  $ P( n_3 ; \delta^- ) $,
as suggested in Eq.\ (\ref{eq:Pc3}), though slightly broader
in the case of $ \delta^- $ due to the uncertainty of the relative phase
for the smaller $ n_3 \approx \bar{n}_3^- $.
The mean-field values $ \bar{n}_3^+ \approx 448 $
and $ \bar{n}_3^- \approx 52 $ are indicated with the vertical dotted lines.

\begin{figure}[t]
\centering
\scalebox{1.2}{
\includegraphics*{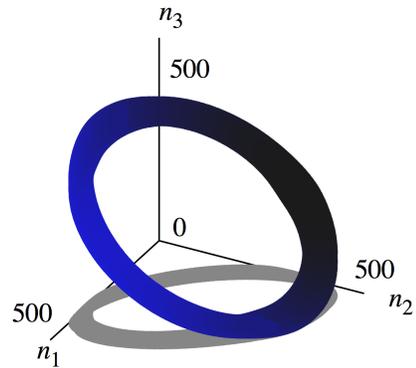}}
\caption{(Color online)
The joint probability $ P (n_1,n_2,n_3) $ is depicted
in the $ (n_1,n_2,n_3) $ space for the Poissonian sources
$ \hat{\Pi}_a \otimes \hat{\Pi}_b $ with $ \bar{N}_a= \bar{N}_b = 500 $.
Here, the points providing significant probabilities
above a certain threshold value $ P_{\rm min} $
(numerically $ 1.6 \times 10^{-7} $) are plotted as dots
to exhibit the peak manifold (blue) along the mean-field trajectory
$ \{ \bar{n}_1 ( \delta ), \bar{n}_2 ( \delta ), \bar{n}_3 ( \delta ) \} $
($ - \pi \leq \delta < \pi $),
together with its projection (gray) on the $ (n_1,n_2) $ plane
along $ \{ \bar{n}_1 ( \delta ), \bar{n}_2 ( \delta ) \} $.
Each of these points is realized as a single-shot interference pattern.}
\label{fig:P3}
\end{figure}

These results of $ P(n_1) $, $ P^\text{c}(n_2|n_1) $
and $ P^\text{c}(n_3|n_1,n_2) $ really indicate
the existence of the peak manifold of $ P({\bf n}_M) $
along the mean-field trajectory.
This is overlooked in Fig. \ref{fig:P12}
for the joint probability $ P(n_1,n_2) $.
The feature of the joint probability $ P (n_1,n_2,n_3) $
of the three detector counts $ (n_1,n_2,n_3) $
is also depicted in Fig. \ref{fig:P3}.
Here, the points providing significant probabilities,
$ P(n_1,n_2,n_3) \geq P_{\rm min} $, are plotted as dots
to exhibit the peak manifold along the mean-field trajectory
$ \{ \bar{n}_1 ( \delta ), \bar{n}_2 ( \delta ), \bar{n}_3 ( \delta ) \} $
($ - \pi \leq \delta < \pi $),
together with its projection on the $ (n_1,n_2) $ plane
along $ \{ \bar{n}_1 ( \delta ), \bar{n}_2 ( \delta ) \} $.
Each of these points is realized as a single-shot interference pattern.
The threshold value of the probability may be chosen suitably
as $ P_{\rm min} \sim 0.1 \bar{P} \sim 0.01 / \bar{n}^2 $
with $ 2 \pi \bar{n} \times \sqrt{\bar{n}} \times \sqrt{\bar{n}}
\times \bar{P} = 1 $ and $ \bar{n}
= ( \langle n_1 \rangle + \langle n_2 \rangle + \langle n_3 \rangle )/3 $;
numerically $ P_{\rm min} = 1.6 \times 10^{-7} $ here.
This calculation of $ P (n_1,n_2,n_3) $ in the $ (n_1,n_2,n_3) $ space
actually provides the simulation of interference experiments
for the three detectors.

\subsection{Sub-Poissonian sources}
\label{sec:na-NB}

\begin{figure}[t]
\centering
\scalebox{1.0}{
\includegraphics*{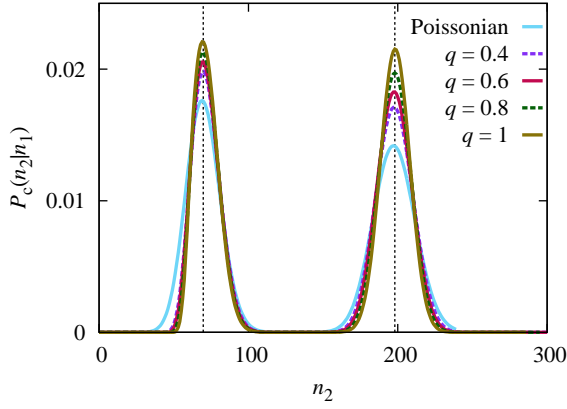}}
\caption{(Color online)
The conditional distributions $ P^\text{c}(n_2|n_1) $ with $ n_1 = 42 $
are plotted for the sub-Poissonian binomial sources
$ \hat{B}_a (q) \otimes \hat{B}_b (q) $
with $ \bar{N}_{a,b} = \bar{N} = 200 $
and some rational values of $ q = \bar{N} / \bar{N}^\prime $
corresponding to the sequence $ | \bar{N} \rangle \langle \bar{N} | (q=1)
\to \hat{B}(q) \to \hat{\Pi} (q \to 0) $.
The mean-field values $ \bar{n}_2^+ \approx 70 $
and $ \bar{n}_2^- \approx 198 $ for $ n_1 = 42 $
are indicated with the vertical dotted lines.
This sequence reproduces equivalently the probability distributions
for the number-state sources $ | \bar{N}/q , \bar{N}/q \rangle $
under the $ q $-scaling with $ R^{(m) \prime} = q R^{(m)} $
and $ \bar{N}^\prime = \bar{N}/q $,
keeping $ R^{(m) \prime} \bar{N}^\prime = R^{(m)} \bar{N} $.}
\label{fig:Pc2-subP}
\end{figure}

As a typical sub-Poissonian case,
consider the binomial state involving the scaling parameter $ q $ (rational)
with a fixed mean photon number $ \bar{N} $ as
\begin{equation}
\hat{B}(q; \bar{N}) = \sum_{N=0}^{\bar{N}/q} B^{\bar{N}/q}_N (q)
| N \rangle \langle N | .
\end{equation}
Then, we have a sequence of the sub-Poissonian states for $ 0 < q \leq 1 $,
\begin{equation}
\hat{B}(q=1) = | \bar{N} \rangle \langle \bar{N} |
\to \hat{B}(q) \to \hat{B}(q \to 0) = \hat{\Pi} .
\label{eq:subPs}
\end{equation}
The conditional distributions $ P^\text{c}(n_2|n_1) $
with $ n_1 = 42 $ are plotted in Fig. \ref{fig:Pc2-subP}
for the binomial sources $ \hat{B}_a (q) \otimes \hat{B}_b (q) $
with $ \bar{N}_{a,b} = \bar{N} = 200 $
and some rational values of $ q = \bar{N} / \bar{N}^\prime $.
Here, the detector matrices $ R^{(m)} $ are taken
the same as in Eq.\ (\ref{eq:R-values})
for the analysis of the Poissonian sources,
while the mean photon number $ \bar{N} $ is somewhat smaller
due to the actual limitation for numerical computation.
This sequence reproduces equivalently the probability distributions
for the number-state sources $ | \bar{N}^\prime , \bar{N}^\prime \rangle
= | \bar{N}/q , \bar{N}/q \rangle $
under the $ q $-scaling in Eq.\ (\ref{eq:R-p-P})
with $ R^{(m) \prime} = q R^{(m)} $
and $ N_{a,b} = \bar{N}^\prime = \bar{N}/q $,
keeping $ R^{(m) \prime} \bar{N}^\prime = R^{(m)} \bar{N} $:
\begin{eqnarray}
P_{\hat{B}(q)}({\bf n}_M ; R^{(m)} \bar{N})
= P_{\hat{N}}({\bf n}_M ; R^{(m) \prime} \bar{N}^\prime ) .
\label{eq:PB-PN}
\end{eqnarray}
Furthermore, we can see in Fig. \ref{fig:Pc2-subP}
that in the limit $ q \to 0 $ the probability distribution
for the binomial sources $ \hat{B}_a (q) \otimes \hat{B}_b (q) $
approaches that for the Poissonian sources
$ \hat{\Pi}_a \otimes \hat{\Pi}_b $.
On the other hand, under the $ q $-scaling
the Poissonian form of statistics is preserved,
and the probability distribution is invariant.
These observations indicate the relation
\begin{eqnarray}
P_{\hat{B}(q \to 0)}({\bf n}_M ; R^{(m)} \bar{N})
&=& P_{\hat{\Pi}}({\bf n}_M ; R^{(m)} \bar{N} )
\nonumber \\
&=& P_{\hat{\Pi}}({\bf n}_M ; R^{(m) \prime} \bar{N}^\prime ) .
\label{eq:PB-PP}
\end{eqnarray}
In the measurement of interference fringes
with continuously distributed $ M $ detectors,
the scaling parameter may be taken as $ q = 1/M $ for the resolution.
Then, by combining Eqs.\ (\ref{eq:PB-PN}) and (\ref{eq:PB-PP})
we find that for $ q \to 0 $ ($ M \gg 1 $)
the number-state sources $ | \bar{N}^\prime , \bar{N}^\prime  \rangle $
and the Poissonian sources $ \hat{\Pi}_a \otimes \hat{\Pi}_b $
with the sufficiently large $ \bar{N}^\prime = \bar{N}/q \to \infty $
provide essentially the same result of interference
with fine spatial resolution as $ R^{(m) \prime} = q R^{(m)} \to 0 $.
By a similar argument, this is also the case for the binomial sources
$ \hat{B}_a (q^{\prime \prime}; \bar{N}^{\prime \prime})
\otimes \hat{B}_b (q^{\prime \prime}; \bar{N}^{\prime \prime}) $
with any rational $ q^{\prime \prime} $
and $ \bar{N}^{\prime \prime} = (q^{\prime \prime}/q) \bar{N} \to \infty $
as $ R^{(m) \prime \prime} = (q/q^{\prime \prime}) R^{(m)} \to 0 $.

\subsection{Super-Poissonian sources}
\label{sec:na-superP}

\begin{figure}[t]
\centering
\scalebox{1.0}{
\includegraphics*{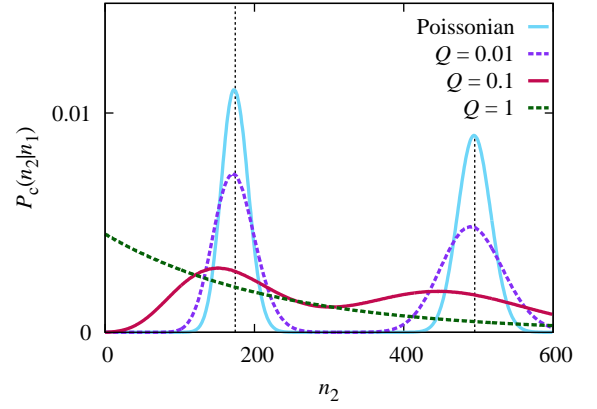}}
\caption{(Color online)
The conditional distributions $ P^\text{c}(n_2|n_1) $ with $ n_1 = 106 $
are plotted for a sequence of super-Poissonian sources
given by $ \mathcal{P}(| \alpha |;Q) $ in Eq.\ (\ref{eq:Palpha})
with $ \bar{N}_a= \bar{N}_b = 500 $ and some values of $ Q $.
The mean-field values $ \bar{n}_2^+ \approx 174 $
and $ \bar{n}_2^- \approx 495 $ for $ n_1 = 106 $
are indicated with the vertical dotted lines.}
\label{fig:Pc2-supP}
\end{figure}

We have also considered a sequence of super-Poissonian states
given by a U(1)-invariant form of coherent-state representation as
\begin{equation}
\mathcal{P}(| \alpha |;Q) \propto (| \alpha |^2/Q \bar{N})^{1/Q-1}
\exp (-| \alpha |^2/ Q \bar{N}) ,
\label{eq:Palpha}
\end{equation}
with the variance $ V $ depending on the parameter $ Q > 0 $,
\begin{equation}
V = \bar{N} + Q \bar{N}^2 .
\end{equation}
The limit $ Q \to 0 $ corresponds to the Poissonian state $ \hat{\Pi} $,
whereas the case $ Q = 1 $ provides the thermal state $ \hat{\Theta} $.
The conditional distributions $ P^\text{c}(n_2|n_1) $ with $ n_1 = 106 $
are plotted in Fig. \ref{fig:Pc2-supP} for some values of $ Q $,
where $ R^{(m)} $ and $ \bar{N}_{a,b} $
are taken the same as in Eqs.\ (\ref{eq:R-values}) and (\ref{eq:N-values})
for the analysis of the Poissonian sources.
The increasing variance $ V $ with $ Q $ broadens the distribution,
as expected, eventually washing out the peak manifold for $ Q \to 1 $.
Hence, the mean-field description is likely invalidated
for super-Poissonian sources with rather broad distributions.

\section{Conclusion}
\label{sec:conclusion}

In conclusion, we have investigated the interference of optical fields
comprehensively under various configurations for sources and detectors.
We have examined the probability distribution of photon detection
to elucidate the quantum theoretical reasoning
for the usual description of interference patterns
with superposition of classical mean fields.
Especially, for interference of two independent mixtures of number states
with Poissonian or sub-Poissonian statistics,
despite lack of intrinsic phases, it has been found
that the joint probability of the photon counts at the detectors
has a distinct peak manifold along the trajectory of the mean-field values
with the varying relative phase.
Then, the interference patterns should mostly be observed shot by shot
as randomly chosen points in the peak manifold,
specifying the values of the relative phase a posteriori.
On the other hand, for super-Poissonian sources the mean-field description
is likely invalidated with rather broad probability distributions.

\begin{acknowledgments}
T. K. was supported by the JSPS Grant No. 22.1355.
\end{acknowledgments}

\appendix

\section{Derivation of the joint probability}
\label{app:derivation}

We present a derivation of the joint probability in Eq.\ (\ref{eq:PM}),
according to Refs. \cite{Kelley1964,Cook1982,Bondurant1985,Vogel2006}.
The probability that $ n_1, \dotsc , n_M $ photoelectrons are emitted
from the respective surfaces $ S_1, \dotsc , S_M $
in the time interval $ T $ is represented \cite{Kelley1964,Vogel2006} as
\begin{equation}
P(n_1, \dotsc , n_M) = \left. \prod_{m=1}^M \frac{1}{n_m!}
\frac{\partial^{n_m}}{\partial z_m^{n_m}}
F(z_1, \dotsc , z_M) \right|_{z_m = -1} ,
\end{equation}
with the generating function
\begin{equation}
F(z_1, \dotsc , z_M) \equiv \sum_{n_1, \dotsc , n_M} P(n_1, \dotsc , n_M)
\prod_{m=1}^M (1 + z_m)^{n_m} .
\end{equation}
This generating function can be expressed in terms of the joint probability
$ w_k(m_1,t_1; \dotsc ; m_k, t_k)dt_1 \dotsm dt_k $
that the $ k $ photoionizations occur, respectively,
at $S_{m_j}$ in the interval $ t_j $ to $ t_j + dt $
($j = 1, \dotsc , k$) \cite{Kelley1964}:
\begin{multline}
F(z_1, \dotsc , z_M) = \sum_{k=0}^\infty \frac{1}{k!}
\sum_{m_1, \dotsc , m_k = 1}^M z_{m_1} \dotsm z_{m_k} \\
\times \int_0^T w_k(m_1, t_1; \dotsc ; m_k, t_k) dt_1 \dotsm dt_k .
\label{eq:Fz}
\end{multline}
A quantum-mechanical expression for $ w_k $ is derived phenomenologically
for the narrow-band field propagating in the $ +z $ direction
\cite{Bondurant1985} as
\begin{multline}
w_k(m_1, t_1; \dotsc ; m_k, t_k) \\
= \eta_{m_1} \dotsm \eta_{m_k} \int_{S_{m_1}} dx_1dy_1 \dotsm
\int_{S_{m_k}} dx_kdy_k \\
\times {\rm Tr} \left[ \mathopen{:}\prod_{l=1}^k
\hat{\psi}^\dagger ({\bf x}_l, t_l) \hat{\psi}({\bf x}_l, t_l)
\mathclose{:} \right] ,
\label{eq:w_k}
\end{multline}
where $ \eta_m $ is the quantum efficiency ($ 0 < \eta_m \leq 1 $),
and the positive-frequency field operator $ \hat{\psi} $
is given in Eq.\ (\ref{eq:psi-op}).
By substituting Eq.\ (\ref{eq:w_k}) into Eq.\ (\ref{eq:Fz})
for the generating function, the expression in Eq.\ (\ref{eq:PM})
for the joint probability $ P(n_1, \dotsc , n_M) $ is obtained
in terms of the photon-flux operators $ \hat{I}_m $ in Eq.\ (\ref{eq:I_m}).
Note here that Eq.\ (\ref{eq:I_m}) coincides with Eq.\ (17)
in Ref. \cite{Cook1982} for the linearly polarized optical field
under the paraxial approximation, which represents the number of photons
that cross the surface $ S_m $ in the time interval $ T $.

\end{document}